# Discovering Physically Interpretable Mathematical Expression for Predicting $CO_2$ Adsorption in Metal–Organic Frameworks via Machine Learning-Symbolic Regression

Yimin Shao,[1,2†] Shengluo Ma,[3†] Shenghong Ju,[3,4*] Yijun Shi,[2] Wei Li[1*]

1. Institute for Materials and Processes, School of Engineering, The University of Edinburgh, Edinburgh EH9 3FB, Scotland, UK
2. Machine Elements Division, Department of Engineering Sciences and Mathematics, Luleå University of Technology, Luleå, 97187, Sweden
3. China-UK Low Carbon College, Shanghai Jiao Tong University, Shanghai 201306, China
4. Key Laboratory for Thermal Science and Power Engineering of Ministry of Education, Department of Engineering Mechanics, Tsinghua University, Beijing 100084, China

†These authors contributed equally to this work.

*Corresponding authors. Email: wli5@ed.ac.uk , jush@tsinghua.edu.cn

## Abstract

This work presents a machine learning–symbolic regression (ML–SR) strategy to develop a physically interpretable formula for predicting low-pressure $CO_2$ adsorption capacity in hypothetical metal–organic frameworks (hMOFs). Four ML models were trained on a small dataset of 1,000 samples, and five key descriptors—largest cavity diameter, pore limiting diameter, void fraction, gravimetric surface area, and number of hydrogen atoms—were identified through SHAP and feature importance analyses. Symbolic regression was then employed to derive a concise adsorption formula, $\boldsymbol{Q = aA}$, where $a$ represents an adsorption baseline (mmol $g^{-1}$) and $\boldsymbol{A}$ is a dimensionless adsorption number incorporating four structural descriptors. We interpret $A$ as the ratio between an adsorption binding force and a diffusion driving force, revealing how pore topology and surface chemistry jointly influence adsorption. Validation against a comprehensive dataset of 137,652 hMOFs demonstrates that this formula achieves over 70% prediction accuracy for 62,448 structures, confirming strong applicability within defined structural and operational ranges. Unlike conventional black-box ML models, the proposed physics-guided expression enables efficient prediction and provides clearer insight into adsorption mechanisms.

## Introduction

Since the industrial revolution, the concentration of carbon dioxide ($CO_2$) in the atmosphere has increased by more than 140 ppm, reaching over 420 ppm as of June 2025 [1]. Carbon capture and storage (CCS) technologies are expected to play a crucial role in mitigating this accumulation, with projections indicating that CCS could reduce 14–20% of total anthropogenic $CO_2$ emissions by 2050 [2 3 4]. In terms of $CO_2$ capture, metal organic frameworks (MOFs) structures, composed of metal oxide clusters connected by organic linkers, have attracted growing attention as promising candidate materials, alongside zeolites and activated carbons [5 6]. However, as the number of MOF variants continues to grow, it has become almost infeasible to experimentally evaluate $CO_2$ adsorption capacities across the entire MOF family. As a more efficient alternative, machine learning (ML) models trained on high-throughput molecular simulation data offer a powerful approach for rapidly assessing $CO_2$ capture performance [7].

In the past five years, numerous studies have applied ML to predict the high-pressure ( > 1 bar) $CO_2$ adsorption capacity in MOFs. Numerous chemical descriptors (e.g., elemental composition, hydrogen bonding, metal content, metallic bonds, single bonds, ring bonds, etc.) and physical–geometric descriptors (e.g., surface area, pore size, density, topological type, etc.) have been employed in ML models to predict the gas adsorption capacities of MOFs [8 9 10]. Various models, including linear regression [11], support vector machines [12], decision trees [7], gradient boosting decision trees [13], random forests [14], eXtreme gradient boosting [15], k-means clustering [16], and artificial neural networks [17], have been employed using different databases (such as hypothetical MOF database [18], CoRE MOF 2019 database [19], CSD MOF [20] and QMOF database [21]). These models have generally achieved prediction accuracies exceeding 80% for $CO_2$ adsorption in MOFs at high pressures [12 15 22].

Considering that practical $CO_2$ capture applications, such as direct air capture [23] and post-combustion capture [24], typically operate at low pressures (≤ 1 bar), the design and development of MOF materials with high $CO_2$ uptake under low-pressure conditions is of particular importance. However, the predictive performance of ML models with existing descriptors decreases significantly under low-pressure conditions [14 25 26]. One possible explanation is that $CO_2$ adsorption in MOFs at high pressure is primarily governed by pore filling, making the geometric characteristics of MOFs closely correlated with adsorption behavior. In contrast, at low pressures, the chemical interactions between gas molecules and the MOF framework may play a more dominant role in determining adsorption performance [25]. So, to predict $CO_2$ adsorption under low-pressure conditions, additional new descriptors (e.g. Henry coefficient [2], atom property–weighted radial distribution function [25]) of MOFs are required to improve the predictive accuracy of ML models

The current research paradigms focus on continuously optimizing ML algorithms or screening novel descriptors of MOFs to enhance the prediction accuracy of $CO_2$ adsorption capacity within existing

MOF databases, often achieving accuracies exceeding 90%. Although these studies have significantly advanced the application of ML in MOF research, particularly contributing to global climate change mitigation and carbon reduction efforts, the rapid development of new MOF materials and the continuous expansion of MOF databases may require frequent retraining of models to maintain prediction accuracy. This research paradigm therefore faces inherent limitations, as it is not highly efficient and can lead to redundant training costs including time, computational, and labor expenses. Therefore, it is essential to explore new research paradigms to address this issue of "repetitive work". Moreover, ML models often operate as a "black box," with limited physical interpretability in their computational processes. The relationships among different descriptors of MOFs and their specific influences on $CO_2$ adsorption capacity remain insufficiently understood.

Then, we propose a new research paradigm that leverages ML models to develop novel mathematical models for gas adsorption. Over the past few decades, numerous mathematical models have been developed to describe gas adsorption isotherms [27] [28], including the Dubinin-Radushkevich model, Langmuir model, Freundlich model, Sips model, Temkin model, and BET model, et al. However, to date, there is no universal mathematical expression capable of accurately predicting the adsorption capacity of a specific gas on a given adsorbent under defined pressure and temperature conditions. Establishing such math models not only aids in revealing the underlying mechanisms and physical meanings of gas adsorption but also enhances model interpretability in accordance with first-principles physics. A model is considered interpretable if the relationship between its inputs and outputs can be logically or mathematically traced in a concise manner. For example, once a mathematical relationship such as Newton's second law, $F = ma$, is derived through ML models, the addition of new data would not invalidate the law itself. In recent years, Symbolic regression (SR) is a rapidly developing subfield of ML that aims to infer symbolic mathematical expressions directly from data [29], demonstrating great potential for scientific discovery. It has been widely applied to uncover fundamental physical laws, such as finding symbolic formulations for Lagrangians [30], Hamiltonians [30,31], Coulomb's law [32], Newton's law of gravitation [31,33], Pauli's spin–magnetic interaction [33], and Schrödinger's wave mechanics [33].

Therefore, this study aims to employ the ML-SR method to find a mathematical formula capable of predicting $CO_2$ adsorption capacity in hMOFs under low-pressure and ambient-temperature conditions, while simultaneously achieving reliable predictive accuracy and more importantly providing meaningful physical insight into gas adsorption mechanisms.

## Results

### *ML model training and feature engineering*

Figure 1 illustrates the overall workflow of the machine learning process used in this study. Initially, a small dataset of 1,000 hMOF samples is employed to train an ML model (Step 1). Feature importance is then analyzed through SHAP interpretation, followed by SR to derive candidate physical formulas (Step 2). These formulas are subsequently validated using a large dataset containing over 137,000 samples (Step 3). Finally, the physically interpretable relationships between descriptors and $CO_2$ adsorption performance are analyzed to provide insight into the underlying structure–property correlations (Step 4).

To predict the $CO_2$ adsorption capacities of hMOFs at 0.5 bar and 298 K, Figure 2 compares the performance and results of 4 ML models: (a) Random Forest (RF), (b) XGBoost (XG), (c) Gradient Boosting Regressor (GBR), and (d) Decision Tree (DT). The vertical axis represents the model-predicted values, while the horizontal axis corresponds to the GCMC-simulated adsorption capacities. The ±1 prediction error range is indicated by the gray dashed area. Although there are noticeable variations between the models, Figure 2(a-d) illustrates how the predicted values from the XG, RF, and GBR models are in good agreement with the ideal fitting line ($y = x$). With a *RMSE* of 0.619 and a five-fold cross-validated $R^2$ of 0.800, the XG model shows the most accurate and concentrated predictions, with the lowest dispersion and highest predictive accuracy. On the other hand, the DT model exhibits a higher *RMSE* of 0.903 and a lower $R^2$ value of 0.574, indicating a higher degree of prediction error and weaker ability to capture intricate nonlinear relationships. The $R^2$ values of the XGBoost, RF, and GBR models for predicting $CO_2$ adsorption capacities of MOFs at low pressure fall within the range of 0.7–0.8, which is consistent with the results [22,34] in the literature.

The distribution of $CO_2$ adsorption capacities in the hMOF dataset is shown in Figure 2(e). The majority of samples have a noticeable uneven distribution and are concentrated in the lower range (0–2). Additionally, violin plots of the normalized coefficient of determination ($tR^2$) and normalized root mean square error (*tRMSE*) derived from numerous cross-validation runs across various models are shown in Figure 2(f). The findings show that the XG model has the best generalization ability, showing the highest concentration and stability for both metrics with comparatively low variance. The DT model, on the other hand, shows greater fluctuations, suggesting that sample imbalance and data noise can affect its performance more.

Figure 3 presents the contribution analysis of 22 structural and chemical descriptors for 4 ML models (RF, XG, GBR, and DT) in predicting $CO_2$ adsorption capacities of hMOFs. Figure 3a shows the mean feature importance values, while Figure 3b illustrates the mean absolute SHAP values. The mean feature importance provides a quantitative assessment of how strongly each model relies on different features during training, and the SHAP analysis reveals the direction of feature influence (positive or negative), as illustrated in Figure S1. Both metrics evaluate the relative importance of each input variable in model prediction, thereby identifying the key factors governing MOF adsorption performance for $CO_2$.

As shown in Figures 3(a-b), although minor variations exist among different models in terms of feature importance values and SHAP values, the overall trends remain highly consistent. The top five influential descriptors, void fraction (voidfrac), pore limiting diameter (pld), specific surface area (s_g), largest cavity diameter (lcd), and the number of hydrogen atoms (H), demonstrate that these structural features play a decisive role in determining MOF adsorption behavior for $CO_2$. Other features, such as the presence of halogens, metal content, and lattice parameters (a, b, c), also show some degree of correlation but contribute considerably less overall. Considering that F and Br are not common chemical elements shared by all MOFs, H was selected as the most suitable chemically relevant descriptor for constructing a generally applicable model. On this basis, we infer that the chemical interaction term in the present formula is primarily associated with the correlation of H atoms. Such chemical interactions may arise from C–H groups on MOF linkers, which can interact more effectively with $CO_2$ through hydrogen bonding and dipole interactions, as supported by previous studies [35] [36] [37] [38]. Therefore, these 5 descriptors can be identified as the key factors governing MOF adsorption behavior for $CO_2$, providing a strong foundation for subsequent symbolic regression analysis.

Figure 4 presents the correlation analysis between various descriptors and $CO_2$ adsorption capacities within the MOF dataset. Figure 4a shows a combined heatmap of the Pearson (blue) and Spearman (green) correlation coefficients, while Figure 4b displays an orange heatmap based on distance correlation. The Pearson correlation coefficient reflects the degree of linear dependence between two continuous variables, ranging from –1 to 1 [39]. Values closer to ±1 indicate a stronger linear relationship, where positive and negative signs represent positive and negative correlations, respectively. The Spearman correlation coefficient, by contrast, is a rank-based nonparametric statistic that measures the strength of a monotonic relationship between variables [40]. Unlike Pearson, it can capture consistent increasing or decreasing trends even when the relationship is nonlinear. Distance correlation measures any type of statistical dependence between two random variables and can capture complex nonlinear or non-monotonic relationships, with values approaching 1 indicating stronger dependence of any kind [41]. Collectively, these three metrics describe linear, nonlinear, monotonic and non-monotonic, relationships among the features, complementing each other in identifying the key structural factors that influence $CO_2$ adsorption in MOFs.

As shown in Figure 4(a), the adsorption capacity exhibits significant positive correlations with surface area (surface_area_m2g), void fraction (void_fraction), largest cavity diameter (lcd), pore limiting diameter (pld), and the number of hydrogen atoms (H), with Pearson and Spearman coefficients approaching 0.8. This indicates that adsorption performance is strongly influenced by the pore structural characteristics of the material, consistent with the physical adsorption mechanism of porous materials, larger surface areas and pore volumes provide more adsorption sites for gas molecules, thereby enhancing adsorption capacity. The distance correlation heatmap in Figure 4(b) further confirms that surface area, void fraction, pld, and lcd show the highest distance correlations with adsorption, reinforcing the dominant role of these parameters in governing adsorption behavior.

## *Symbolic regression*

The SR method operates by combining evolutionary operators, such as addition, subtraction, multiplication, division, exponentiation, and logarithmic functions, to automatically search for the optimal analytical expression that minimizes prediction error while maintaining mathematical simplicity [29]. Unlike conventional black-box ML models, SR produces explicit mathematical equations with clear physical meaning. In this SR process, the input descriptors consist of 5 parameters closely related to the structural and chemical characteristics of MOFs: $X_0$ = lcd / 10; $X_1$ = pld / 10; $X_2$ = void fraction; $X_3$ = surface area / 1000; $X_4$ = number_H. During data preprocessing, certain variables were divided by fixed scaling constants (e.g., 10 or 1000) to achieve non-dimensionalization. This procedure converts variables with different physical units and magnitudes into comparable, unitless quantities, preventing features with large numerical values from dominating the regression optimization process. Such scaling enhances both the numerical stability and the physical interpretability of the derived formulas. By applying this non-dimensionalization approach, the model can identify genuine interdependencies among variables on a consistent scale, rather than being biased by differences in dimensional units.

The density distribution plot in Figure 5(a) illustrates the relationship between the fitted Pearson correlation coefficient (*r*) and formula complexity (Formula Length) for 2,641 candidate expressions shown in Table S3. In statistical terms, *r* represents the correlation coefficient used to measure the strength of the linear relationship between two datasets. As observed, *r* increases with formula complexity, but this trend saturates when the complexity reaches approximately 15. This indicates that a moderately complex analytical expression is sufficient to achieve a high *r* value, and further increasing the complexity does not lead to a significant improvement. 8 high-performing formulas, marked by red points, were selected from this equilibrium region [42 43], balancing predictive accuracy and model simplicity for subsequent analyses. Since the optimization objectives are to maximize the absolute Pearson correlation while minimizing complexity, the Pareto front is therefore located along the upper-left boundary of the distribution shown in the Figure 5(a). This concept has been widely applied in multi-objective optimization problems in machine learning [44 43 45].

Regarding formula selection process in Figure S3, we firstly screened the 2,641 preliminary symbolic regression equations by retaining those with an absolute Pearson correlation coefficient higher than 0.75 and a formula length shorter than 16. During this process, we also attempted to preserve formulas located near the Pareto front as much as possible. Subsequently, based on our preliminary interpretation of mathematical operators shown in Table S3, addition was considered to represent superposition effects, subtraction to indicate differences, multiplication to reflect amplification or coupling, and division to denote density or normalization. We then examined whether the five descriptors were expressed within consistent dimensional units. It was found that only $X_0$ and $X_1$ share the same dimensional unit (Å), whereas the other descriptors possess their own distinct dimensions. Therefore, when descriptors such as $X_2$, $X_3$, $X_4$, $X_0$, and $X_1$ appear in addition or subtraction operations, their physical meaning cannot be reasonably interpreted under our current understanding. As a result, we chose to discard such formulas, even when they exhibited relatively good predictive accuracy. Because these 2,641 equations were also generated through the random combination process of the genetic algorithm in symbolic regression, the selection of a final formula should not rely solely on predictive accuracy; another equally important consideration is interpretability. However, physical or chemical interpretability often depends strongly on the researcher's own experience in the relevant field, as well as original scientific insight.

We did not prioritize $X_1$ for the following reasons. (1) As shown in Figure 4a, the blue heatmap represents Pearson correlation. From both linear and nonlinear perspectives, we found that $X_1$ (PLD) exhibits the strongest nonlinearity among all descriptors. However, our target in this work was to identify relatively simple linear-form analytical equations. This mismatch between the intrinsic nonlinear behavior of PLD and the linear structure sought in the symbolic regression search reduces the probability of $X_1$ being selected during formula generation, as reflected in Figures S4 (a-b). (2) As shown in Figure S4(c), many formulas containing $X_1$ were not located close to the Pareto front. Consequently, among the 8 selected candidate formulas, only one contained $X_1$. (3) Most importantly, based on our current understanding of adsorption mechanisms, the formulas generated by symbolic regression that contain $X_1$ are still difficult to interpret in terms of physical reasonableness and mechanistic meaning.

Figure 5(b–i) illustrates the predictive performance of eight high-performing analytical expressions derived from symbolic regression on the test samples. Most of the scatter points are distributed closely along the linear fitting line, and the Pearson correlation coefficients ($r$) of the formulas range from 0.69 to 0.81, indicating a strong linear consistency between the predicted values and the GCMC-simulated data. Among them, the formulas shown in Figures 5(b) and 5(d), $(X_2\, X_3\, X_4^{1/4}) / [X_0\, (1 + X_3^2)]$ and $((X_2 - 0.147\, X_3)\, (X_3\, X_4\, )^{1/4}) / X_0$, exhibit the highest correlations with ($r$ = 0.7805) and ($r$ = 0.8059), respectively. SR automatically identifies and combines these parameters, successfully constructing function forms consistent with physical principles. More importantly, it reveals explicit mathematical relationships between MOF structural descriptors and $CO_2$ adsorption performance. The subsequent analysis will further evaluate the validity of these equations using the large-scale dataset.

Figure 6 presents the generalization performance of 8 SR-derived formulas on a large-scale hMOF dataset including approximately 137,000 samples. To improve fitting accuracy, a proportional coefficient ($a_i$) was introduced for each equation to compensate for systematic deviations of the symbolic expressions across different structural scales. Figure 6(a–h) show scatter plots comparing the each formula-predicted and GCMC-computed $CO_2$ adsorption capacities on the large validation set.

The red solid line denotes the ideal fitting line (y = x), while the color gradient of the scatter points (from light yellow to dark green) represents the range of $CO_2$ adsorption capacities (mmol $g^{-1}$). Each subplot is labeled with the corresponding equation and its coefficient of determination ($R^2$), which quantifies the goodness of fit between the symbolic prediction and the simulation data.

Overall, the SR–derived formulas exhibit distinct performance differences when applied to the large-scale dataset. The formula in Figure 6(c) achieves the highest $R^2$ of 0.634, indicating its strong capability to capture the nonlinear relationship between MOF structural features and $CO_2$ adsorption capacity. Formulas in Figure 6(a, b, d, g) show moderate fitting accuracy, with $R^2$ values ranging from approximately 0.50 to 0.60, whereas formula in Figure 6(f) yields an $R^2$ of only 0.003, suggesting almost no correlation and poor generalizability in the complex sample space. Although some formulas perform well for specific subsets of data, the overall generalization ability varies markedly with formula complexity and the degree of variable interaction across the entire dataset of 137,000 samples.

It is worth noting that, among these 8 candidates in Table S2, only 3 formulas $(X_2\ X_3\ X_4^{1/4})\ /\ [X_0\ (1 + X_3^2)]$ , $((X_2 - 0.147\ X_3)\ (X_3\ X_4\ )^{1/4})\ /\ X_0)$ , $X_0\sqrt{X_4} + X_2\ /\ 0.059X_0(X_3 - \log(X_3))$ predicted more than 60,000 MOFs with an absolute relative error below 30%, and were therefore validated as effective candidate formulas. These results indicate that symbolic regression can not only effectively capture the dominant structural features learned from small-sample training, but also achieve relatively high predictive accuracy on large-scale datasets containing approximately 20,000 to 60,000 entries, while maintaining good interpretability.

As shown in Figure S2, although the empirical formulas proposed in our work yields an overall $R^2$ value of 0.596 for the 137,000-entry MOFX-DB database, its performance improves substantially within practical error ranges. Specifically, for 78,039 hMOFs with an absolute relative error below 40%, the $R^2$ reaches 0.844. Furthermore, for 62,448 hMOFs with an absolute relative error below 30%, the $R^2$ further increases to 0.907. We honestly acknowledge that the resulting predictive accuracy cannot match those models incorporating newly engineered descriptors reported in previous studies [46] [2] [25]. However, for the objectives of the present work, we believe that this strategy successfully achieved our intended research goals.

We believe that the moderate predictive performance observed on the large-scale dataset is likely attributable to the limitations of the currently available descriptors. Previous studies [22] [2] [46] have demonstrated that, under different pressure conditions, the predictive accuracy ($R^2$) of models with the same descriptors can vary significantly. In addition, the incorporation of new key descriptors would likely help improve the predictive accuracy ($R^2$) of models for $CO_2$ adsorption capacity under low-pressure conditions.

## ***Physical insights from an interpretable formula***

Therefore, we selected the expression $a_1$ ($X_2$ $X_3$ $X_4^{1/4}$) / [$X_0$ (1 + $X_3^2$)] as the preferred formula, since it offers clearer physical interpretability, even though the alternative expression $a_3$ (($X_2$ - 0.147 $X_3$) ($X_3$ $X_4$ )$^{1/4}$) / $X_0$ yields a slightly higher $R^2$ value. The reason we did not select the formula, $a_3$ (($X_2$ - 0.147 $X_3$) ($X_3$ $X_4$ )$^{1/4}$) / $X_0$, is, frankly speaking, that based on our current understanding, we were unable to provide a reasonable physical interpretation for it. In particular, the term $X_2$ - 0.147 $X_3$, representing void fraction minus surface area, could not be meaningfully rationalized from a physical perspective. Precisely because the SR method itself does not possess an independent capability to interpret the physical meaning of the generated formulas.

Based on Equation in Figure 6a, we propose the following empirical formula for estimating gas adsorption capacity under low-pressure and room-temperature conditions:

$$\boldsymbol{Q} = \boldsymbol{a}\frac{\boldsymbol{X_2 X_3 X_4^{1/4}}}{\boldsymbol{X_0(1 + X_3^2)}} \tag{1}$$

Where, *a* represents the adsorption baseline constant. We propose that this constant corresponds to the baseline $CO_2$ adsorption capacity of a MOF adsorbent at 0.5 bar and 298 K, with a value of 3.3 mmol $g^{-1}$. We assume that this adsorption baseline constant may depend on the gas species, temperature, and pressure; however, this remains a reasonable hypothesis at present and requires further investigation. The scaled parameters are defined as follows: $X_0 = lcd / b_0$, where $b_0 = 10$ Å and lcd is the largest cavity diameter (Å); $X_1 = pld / b_1$, where $b_1 = 10$ Å and pld is the pore limiting diameter (Å); $X_2$ = *void fraction*, a dimensionless structural parameter; $X_3 = Sa / b_3$, where $b_3 = 1000$ $m^2$ $g^{-1}$ and *Sa* denotes the surface area ($m^2$ $g^{-1}$); and $X_4$ = *H number*, a dimensionless quantity. This equation provides an interpretable and physically consistent representation of the relationship between MOF structural parameters and $CO_2$ adsorption capacity.

Based on the above equation, we introduce a dimensionless characteristic number to describe the adsorption capacity of MOFs, denoted as the adsorption number (***A***):

$$\boldsymbol{A} = \frac{\boldsymbol{X_2 X_3 X_4^{1/4}}}{\boldsymbol{X_0(1 + X_3^2)}} \tag{2}$$

Dimensionless numbers [47] are widely used in fluid mechanics, such as the Reynolds number (*Re*), Nusselt number (*Nu*), and Froude number (*Fr*). These quantities serve as bridges for cross-scale and cross-system comparisons and analyses of complex physical processes. They help unify underlying laws, highlight essential mechanisms, and reduce both experimental and computational costs. In a similar manner, we expect the adsorption characteristic number ***A*** to possess these same advantageous properties, providing a generalized and physically interpretable descriptor for evaluating MOF adsorption behavior.

We propose that the term ($X_2\ X_3\ X_4^{1/4}$) empirically represents the adsorption binding force, which is closely related to the surface chemistry of the MOF. This relationship is reflected in the equation through $X_4$, which denotes the number of hydrogen atoms, an indicator of surface chemical activity. Firstly, the term ($X_2\ X_3\ X_4^{1/4}$) represents that the adsorption binding force arises not only from the contribution of chemical interactions (as represented by $X_4$), but also from interactions at the physical structural level (as represented by $X_2$ and $X_3$). These two contributions are mutually synergistic and coupled. This interpretation is consistent with the fundamental mechanism of gas adsorption on adsorbent materials [48]. Next, we would like to explain the mathematical meaning of multiplication in this context. Multiplication can represent scaling (amplification/reduction) or coupling. For example, in y=ax, if a>1, the effect is amplified, whereas if 0<a<1, the effect is reduced. Accordingly, if $X_2\ X_3 > 1$, the contribution of $X_4$ is amplified; if $0 < X_2\ X_3 < 1$, the contribution of $X_4$ is diminished. The term $X_4^{1/4}$ is a power-law function, introduced to reflect that once the variable reaches a certain magnitude, its growth rate gradually slows while remaining monotonically increasing, as shown in Figure 7d. The selection of this exponent was identified by the SR algorithm through extracting such patterns from the large-scale dataset.

Furthermore, we define $X_0\ (1 + X_3^2)$ as an empirical expression describing the diffusion driving force. The diffusion driving force is strongly correlated with the physical and topological characteristics of the MOF adsorbent, as captured by the presence of the largest cavity diameter (lcd, represented by $X_0$) and the surface area ($X_3$) in the formula. Shima et al. [49] reported through pulsed-field gradient nuclear magnetic resonance (PFG NMR) experiments that gas diffusion in porous materials is strongly governed by pore aperture size, network connectivity, tortuous transport pathways, and framework flexibility, all of which determine molecular accessibility, hopping barriers, and intracrystalline mobility. For this reason, we selected a pore-size-related descriptor, namely largest cavity diameter.

The constant “1” in the denominator is treated as a physical factor constant. We assume that, in future studies, this constant may be correlated with the gas diffusion coefficient; at present, we propose this only as a reasonable speculation for readers to consider. For the current dataset, this physical factor is set to 1. Since the diffusion driving force is inversely related to confinement effects, the reciprocal of this term (i.e., 1/diffusion driving force) may be interpreted as diffusion resistance.

Therefore, the adsorption characteristic number $\boldsymbol{A}$ can be viewed as the ratio between adsorption binding force and the diffusion driving force. A larger $\boldsymbol{A}$ value indicates a stronger adsorption capability of the MOF adsorbent. When the diffusion driving force increases, corresponding to weaker confinement within the pores, the value of $\boldsymbol{A}$ decreases, suggesting reduced adsorption performance. In contrast, stronger adsorption binding strength leads to a higher $\boldsymbol{A}$ value, implying enhanced $CO_2$ uptake capacity of the MOF material.

Figure 7 presents the correlation analysis between $\boldsymbol{A}$ and the main feature descriptors (e.g., lcd, void fraction, surface area, and number of hydrogen atoms), including both single-factor and two-factor analyses. Figure 7(a–d) illustrate the variation of ***A*** with respect to lcd ($X_0$), void fraction ($X_2$), surface area ($X_3$), and number of hydrogen atoms ($X_4$), respectively. Figure 7(e–h) further illustrate the synergistic effects of paired structural features on ***A***. All results are derived from the empirical expression $(X_2 X_3 X_4^{1/4}) / [X_0 (1 + X_3^2)]$, whose explicit mathematical form enables quantitative interpretation of how individual and combined structural parameters influence the adsorption performance of MOFs.

As shown by the single-factor analysis in Figure 7(a–d), the ***A*** exhibits an inverse relationship with the largest cavity diameter. As the pore size increases, ***A*** decreases significantly, indicating that excessively large pores are unfavorable for the effective retention of $CO_2$ molecules. Larger pore sizes weaken the confinement effect of the adsorbent, which is consistent with the experimental observations and mechanistic explanations reported by Snurr et al. [50], who revealed that smaller pores enhance $CO_2$ adsorption at low pressures. Micropores (diameter < 20 Å) are particularly effective for capturing small gas molecules such as $N_2$ and $CO_2$, whereas mesopores (20–500 Å) are more suitable for adsorbing larger organic molecules and complex pollutants.

Figure 7(e) shows that ***A*** reaches its maximum at small lcd values and a specific surface area of approximately 1000 $m^2 g^{-1}$, indicating the existence of an optimal matching region between pore size and surface area. Figure 7(f) demonstrates that adsorption performance significantly improves when lcd is small and the void fraction is high, suggesting that compact frameworks with high porosity are more favorable for $CO_2$ capture under low-pressure conditions. In Figure 7(g), a clear synergistic effect between surface area and porosity can be observed: ***A*** attains its maximum at high void fractions and a surface area of around 1000 $m^2 g^{-1}$, again implying the existence of an optimal matching region between these two parameters. This finding is consistent with many experimental observations in Yu et al.'s study [48], which reported that excessively large surface areas and pore sizes lead to reduction the affinity of $CO_2$ molecules toward the pore surface. Figure 7(h) further indicates that an optimal matching region also exists between high hydrogen content and a surface area of approximately 1000 $m^2/g$, reinforcing the importance of balanced structural and chemical features for maximizing $CO_2$ adsorption. Therefore, the explicit equation $a\ (X_2 X_3 X_4^{1/4}) / [X_0 (1 + X_3^2)]$ provides a theoretical foundation and a computable adsorption model for subsequent MOF structural optimization and rapid performance screening.

## *Potential application from an interpretable formula*

Figure 8a compares the $CO_2$ adsorption capacities of seven common MOFs at 0.5 bar predicted by the empirical formula, obtained through symbolic regression, with the experimental data (Exp). The selected MOFs include Cu-BTC (HKUST-1), IRMOF-1 (MOF-5), MIL-101(Cr), NU-1000, MIL-100(Cr), UiO-66, and ZIF-8, covering a range of metal centers and topological types, and representing diverse structural characteristics from rigid to flexible frameworks and from microporous to mesoporous materials. From the overall trend, the empirical equation derived from symbolic regression reproduces the adsorption behavior of most MOFs reasonably well. For structures such as HKUST-1, MOF-5, MIL-100(Cr), MIL-101(Cr), MIL-101(Cr)-$NH_2$ and NU-1000, the deviation between the predicted values and experimental results is within 10%, as shown in Table 1, demonstrating good generalization ability and physical interpretability. For these six MOFs, the linear regression coefficient of determination between experimental and predicted values reaches $R^2 = 0.977$ in Figure 8a.

However, for the flexible MOFs ZIF-8 and UiO-66, the predicted adsorption capacities are significantly lower than the experimental values, indicating that the empirical equation still has certain limitations in describing structurally dynamic systems. This discrepancy may be attributed to the fact that the input descriptors of the symbolic regression model are primarily derived from static crystal structures (e.g., void fraction, surface area, and pore size), while the dynamic "breathing effect" of the framework or molecular sieving behavior is not explicitly considered [48].

The empirical formula proposed in this work can also provide a quantitative strategy for addressing how MOFs may be modified to enhance their $CO_2$ adsorption capacity. Previous approaches for improving $CO_2$ uptake in MOFs have largely relied on qualitative conclusions derived from past experimental experience, such as introducing amino functional groups into the framework. However, the subtle synergistic relationships among LCD, void fraction, and surface area have not been systematically resolved in most previous studies, which substantially increases the randomness and trial-and-error nature of experimental design.

As illustrated in Figure 8d, our empirical formula can be used to estimate the number of H atoms associated with the functional groups introduced into a target modified MOF. The total number of H atoms can then be used as the $X_4$ parameter, followed by determining $X_0$. Subsequently, reasonable ranges of $X_0$, $X_2$, and $X_3$ can be assigned to calculate the final $Q$ value ($CO_2$ adsorption capacity), thereby assisting in the rational modification and design of MOF structures.

For example, as illustrated in Figures 8b and 8c, all H atoms in MIL-101(Cr) originate from the benzene rings of the organic linkers. Assuming that one H atom on each benzene ring is replaced by an $NH_2$ functional group, the total number of H atoms in pristine MIL-101(Cr), obtained from its CIF file [51], is 3264. On this basis, the total number of H atoms in the modified MIL-101(Cr) can be theoretically estimated as 4080. After determining $X_4$ (the number of H atoms), a 3-dimensional correlation map can be established between the three parameters $X_0$, $X_2$, and $X_3$ and the $CO_2$ adsorption capacity. Subsequently, $X_0$(LCD) may be selected as a second fixed variable. As shown in Figure 8b, a value of $X_0 = 25$ Å was chosen based on reported experimental studies of $NH_2$-functionalized MIL-101(Cr) [52]. As shown in Figure 8c, by comparing the position of the green point representing pristine MIL-101(Cr) with the red point corresponding to the experimentally reported value in the literature, the optimal functionalization and optimization pathway can be identified from the correlation map. More importantly, Figure 8c suggests the possibility of achieving $CO_2$ adsorption capacities even higher than the reported experimental value (2.5 mmol $g^{-1}$).

Such a modification strategy greatly enhances the purposefulness, feasibility, visual interpretability, and exploratory potential of future experiments. Moreover, it can accelerate the optimization and design of high-performance adsorbent materials, promote the large-scale production of MOFs, and contribute to the realization of carbon neutrality goals.

## Discussion

This study proposes a novel research strategy that integrates machine learning with SR to develop physically interpretable mathematical expressions for predicting the $CO_2$ adsorption capacity of large datasets of hMOFs under low-pressure conditions. This approach enables the automatic extraction of physically meaningful equations from complex data, providing a pathway that effectively bridges data-driven and mechanism-driven methodologies. It offers an important theoretical foundation for high-throughput materials screening and structural optimization. The main findings of this work are as follows:

By training 4 machine learning models (RF, XGB, GB, and DT) on a small dataset, and using SHAP values and feature importance analysis, 5 key structural descriptors were identified ($X_0 = lcd$, $X_1 = pld$, $X_2 = void\ fraction$, $X_3 = surface\ area$, $X_4 = number\ of\ hydrogen\ atoms$). Through SR, a candidate formula $(X_2 X_3 X_4^{1/4}) / [X_0 (1 + X_3^2)]$ was derived, which exhibited certain level of predictive capability on the large-scale dataset. In particular, for 62,448 MOFs, the relative absolute error between the predicted and reference values was below 30%. Moreover, the linear regression coefficient of determination between the predicted values and reference data for these 62,448 MOFs reached R^2=0.907. So, the proposed equation significantly reduced computational cost and improved efficiency.

We propose the formula $Q = aA$, where $a$ represents the adsorption baseline (mmol $g^{-1}$) and $\boldsymbol{A}$ is a dimensionless adsorption number that incorporates four structural descriptors, defined as $\boldsymbol{A} = (X_2 X_3 X_4^{1/4}) / [X_0 (1 + X_3^2)]$. This expression provides a physically interpretable relationship between the descriptors and the $CO_2$ adsorption performance of MOFs. The dimensionless parameter $\boldsymbol{A}$ represents the ratio of the adsorption binding force $(X_2 X_3 X_4^{1/4})$ to the diffusion driving force $[X_0 (1 + X_3^2)]$, determining the overall strength of $CO_2$ adsorption in MOF adsorbents. Moreover, the dimensionless adsorption number $\boldsymbol{A}$ serves as a bridge for cross-scale and cross-system comparison of complex gas adsorption processes. It enables the unification of governing trends, highlights essential mechanisms, and helps reduce both experimental and computational costs. The adsorption binding force is strongly related to the surface chemistry of MOFs, which is reflected in the inclusion of $X_4$ as a key surface chemical element factor. The diffusion driving force, in contrast, is closely associated with the physical topology of the MOF structure. As the diffusion driving force increases, the confinement effect of the MOF topology becomes weaker, leading to a decrease in the $A$ value. This indicates that the diffusion driving force is negatively correlated with the topological confinement strength of the MOF framework.

A quantitative single- and two-factor correlation analysis between different structural descriptors and the adsorption number $\boldsymbol{A}$ reveals that the void fraction and the number of hydrogen atoms are positively correlated with $\boldsymbol{A}$, whereas lcd shows a negative correlation. Moreover, the surface area exhibits an optimal matching range with $\boldsymbol{A}$. Unlike conventional machine learning models, the proposed formula not only enables efficient prediction but also provides a theoretical foundation for subsequent MOF structure optimization.

We acknowledge that, compared with the full range of experimentally reported MOFs, the chemical diversity of MOFX, particularly in terms of metal centers and inorganic nodes, remains limited. Therefore, the present formula should not be interpreted as a universally complete expression applicable to all MOFs, but rather as a data-driven and interpretable relationship that is valid within a certain scope. Its applicability to MOFs with richer metal chemistry will require future validation using more chemically diverse databases and experimental datasets. We also recognize that the current equation has limitations, especially in capturing surface chemical interactions, and that this aspect warrants further investigation.

In addition, we acknowledge that human involvement in the interpretation and selection of symbolic regression equations inevitably introduces a degree of subjectivity and contingency. This consideration was, in fact, one of the motivations for publishing the present work and openly sharing all 2,641 candidate formulas. By making these results publicly available, we hope to encourage broader participation from researchers interested in this emerging research paradigm. As more scholars contribute to the interpretation, evaluation, and refinement of these candidate equations, the influence of individual bias can be progressively reduced, ultimately leading to a broader and more robust scientific consensus.

## Methods

### *Data Collection and Feature Preparation*

The MOFX DB database [18], containing GCMC simulation results for 137,652 hMOFs at 0.5 bar and 298 K, serves as the primary data source. Structural descriptors include pore limiting diameter (PLD), largest cavity diameter (LCD), void fraction, density, gravimetric surface area, accessible volume, chemical composition, and force field parameters. A subset of 1,000 MOFs was randomly sampled as a representative training set, consistent with prior SR studies where smaller datasets facilitate interpretable formula discovery. We agree that the relatively short GCMC simulations used in constructing the database may introduce statistical uncertainty, which indeed increases the difficulty of symbolic regression. In this sense, even in the presence of such noise, symbolic regression generally tends to capture the dominant global correlations rather than overfitting local fluctuations, which can partially mitigate the influence of stochastic noise. Moreover, since we are currently unable to alter the conditions of the original database, we sought to address potential statistical noise and bias in the samples by first performing symbolic regression on a small subset of 1,000 samples to identify candidate equations, and then evaluating these candidate formulas on the large-scale dataset containing more than 137,000 entries. This two-step strategy provides an additional assessment of the stability and robustness of the derived formulas.

We selected the MOFX database mainly based on the following considerations: (1) The database contains more than 137,000 hMOF structures, providing a sufficiently large search space and statistical foundation for symbolic regression. (2) The database simultaneously provides structural descriptors and adsorption data obtained under a unified computational workflow, thereby avoiding methodological inconsistencies among data collected from different literature sources. (3) Its standardized features (e.g., pore size, surface area, void fraction, and compositional information) are highly suitable for interpretable formula discovery.

***Construction of ML Models***

Four ML models, Random Forest (RF) [53,54], extreme Gradient Boosting (XGB) [55], Gradient Boosting (GB) Regression [56], and Decision Tree (DT) [57], were implemented using the Scikit-learn library [58] and evaluated through five-fold cross-validation. The coefficient of determination ($R^2$) and the root mean square error (*RMSE*), both obtained from the same module, were employed to evaluate the ML model's performance. Each model was independently trained 200 times to ensure robustness and reproducibility of the results. For each independent training, hyperparameter tuning was performed using the Bayesian Optimization package [59], a global optimization framework, to maximize predictive accuracy as measured by $R^2$. In this procedure, the Gaussian process and acquisition function were first initialized with ten randomly selected parameter sets, followed by 100 optimization iterations to identify the optimal hyperparameters for each model. A detailed list of the input features and their corresponding descriptions is provided in Table S1.

***Feature Analysis and Selection***

To elucidate the relationship between the descriptors and $CO_2$ adsorption capacity, the SHAP [60] toolkit was employed to quantify feature importance across the 4 machine learning models. SHAP analysis, which is grounded in game-theoretic Shapley values, enables interpretation of the individual contribution of each feature to the model predictions. Swarm plots, generated from the best-performing model within each type of ML algorithm, were used to visualize the distribution of SHAP values for individual features. Given that SHAP results depend on model training, the mean SHAP contribution values derived from the top 25 trained models were used as a reference for comparing feature importance during feature selection. This approach helps to minimize the potential influence of model variability and feature perturbations. Detailed descriptions of the features are provided in Table S1.

***Feature Engineering via Symbolic Regression***

The SR [42,44] formula construction strategy was adopted in the development of simplified screening descriptors and implemented using the gplearn package [61]. Considering that genetic optimization in SR is not well-suited for high-dimensional datasets, only five key physical descriptors were employed as input variables: largest cavity diameter (LCD), pore limiting diameter (PLD), void fraction (VoidFrac), gravimetric surface area (Sa), and the number of hydrogen atoms (H) in the crystal's chemical composition. Detailed definitions of these descriptors are provided in Table 2.

The Pearson correlation coefficient was adopted as the fitness function in the SR analysis of the smaller training dataset (1,000 MOFs), with the goal of generating new descriptors that exhibit stronger linear correlations to serve as screening criteria. Weng et al. [45] reported that using a smaller dataset can facilitate SR analysis, allowing for the identification of simple yet meaningful descriptors and the construction of mathematical formulas that best fit the certain dataset. A grid search procedure was further applied, using the hyperparameters and evaluation metrics summarized in Table 3, to identify the mathematical expressions. The column labelled "Combination" in Table 3 represents the number of tested hyperparameter settings in the grid-search space during symbolic regression optimization. Eight formulas located on the Pareto front were obtained through a bivariate density distribution approach, along with the optimal and most concise formulas derived from the initial set of five variables. The Pareto front refers to a class of statistical methods that identify and approximate the Pareto-optimal boundary by estimating the joint density (or cumulative distribution) in a bi-objective space and using the boundary of that density distribution. The core idea is that when simultaneously optimizing multiple conflicting objectives (for example, maximizing accuracy while minimizing complexity), a single global optimum usually does not exist. Instead, there is a set of non-dominated trade-off solutions, and the boundary formed by these solutions is called the Pareto front.

During data preprocessing, certain variables were divided by fixed scaling constants (e.g., 10 or 1000) to achieve non-dimensionalization. This procedure converts quantities with differing magnitudes and physical units into comparable, unitless forms, thereby preventing variables with large numerical ranges from disproportionately influencing model optimization. Such non-dimensionalization enhances the numerical stability and physical interpretability of the resulting formulas, allowing the model to discern genuine interdependencies among variables rather than artifacts introduced by differences in dimensional scales.

According to the symbolic regression hyperparameter settings of the gplearn software listed in Table

3, the total cumulative evolutionary search scale and the total number of generated formula results can be estimated as follows:

(1) Total symbolic regression search evaluations = generations × population size × operator-probability combinations × random seeds × initialization depths;

(2) Total symbolic regression configurations = operator-probability combinations × random seeds × initialization depths.

Accordingly, the total number of symbolic regression search evaluations exceeded

$$200 \times 1000 \times 746 \times 5 \times 4 = 2{,}984{,}000{,}000$$

while the total number of generated formula outputs was

$$746 \times 5 \times 4 = 14{,}920$$

It should be noted that, even under different hyperparameter settings, symbolic regression may still produce identical or highly similar expressions. To eliminate the influence of such redundant results on statistical analysis, this study converted each evolved symbolic regression formula, based on its expression-tree structure, into a standardized S-expression (Symbolic Expression) tree-text representation. This string representation was then treated as the unique ID of the formula and used to determine whether two formulas were identical, thereby enabling deduplication of the search results. By removing repeated local optima, potential bias in formula frequency statistics, performance evaluation, and interpretability analysis can be effectively avoided, thereby improving the accuracy and reliability of subsequent analyses. As a result, we ultimately obtained 2,641 unique symbolic regression formulas.

## Data Availability

The datasets generated and/or analyzed during the current study are available from the corresponding author upon reasonable request. Detailed descriptions can be found in the Methods Section and Supplementary Information.

## Code Availability

The scripts used for ML, SHAP analysis, symbolic regression and relevant data analysis are available from the corresponding author upon reasonable request.

## Acknowledgements

W. Li and Y. Shao acknowledge the financial support from the European Commission H2020 Marie S Curie Research, Innovation Staff Exchange (RISE) award (Grant No. 871998) and 2023-24 SJTU-UoE Joint Seed Fund. Additionally, S. Ju and S. Ma acknowledge the financial support from the 2023-24 SJTU-UoE Joint Seed Fund, the Shanghai International Science and Technology Collaboration Project (No. 24160712600), and the Shanghai Municipal Education Commission Research Project (No. 2024AIZD012). S. Ma and Y. Shao also acknowledge the technical support from the Center for High-Performance Computing at Shanghai Jiao Tong University to perform computations on the π 2.0 cluster.

## Author contributions

Y. Shao and S. Ma designed the research, conducted the computations, analyzed the data and interpreted the results. Y. Shao and S. Ma jointly drafted the manuscript. W. Li and S. Ju conceptualized, supervised the research and secured fundings. Y. Shao, S. Ma, W. Li, S. Ju and Y. Shi contributed to the discussions and provided feedback on the manuscript.

## Competing Interests

The authors declare no competing financial or non-financial interests.

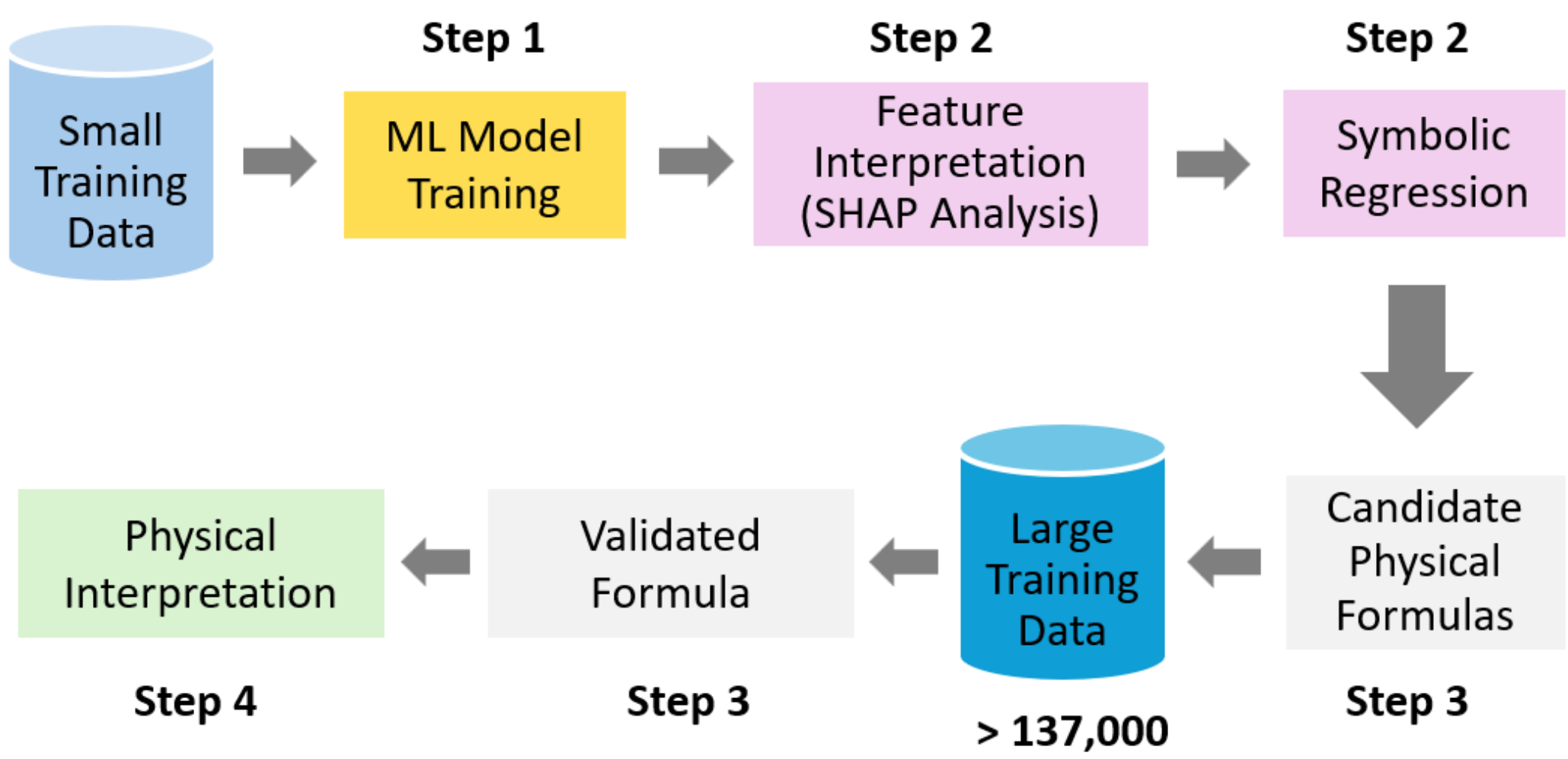

Figure 1. Workflow diagram of the machine learning (ML) process. This figure mainly illustrates the 4 key steps employed in this study to obtain interpretable equations. Firstly, a subset containing 1,000 hMOF samples was used to train ML models, and the key descriptors were identified through SHAP analysis. Subsequently, symbolic regression was applied to generate candidate equations with physical interpretability. The resulting formulas were then validated using a large dataset containing more than 137,000 samples.

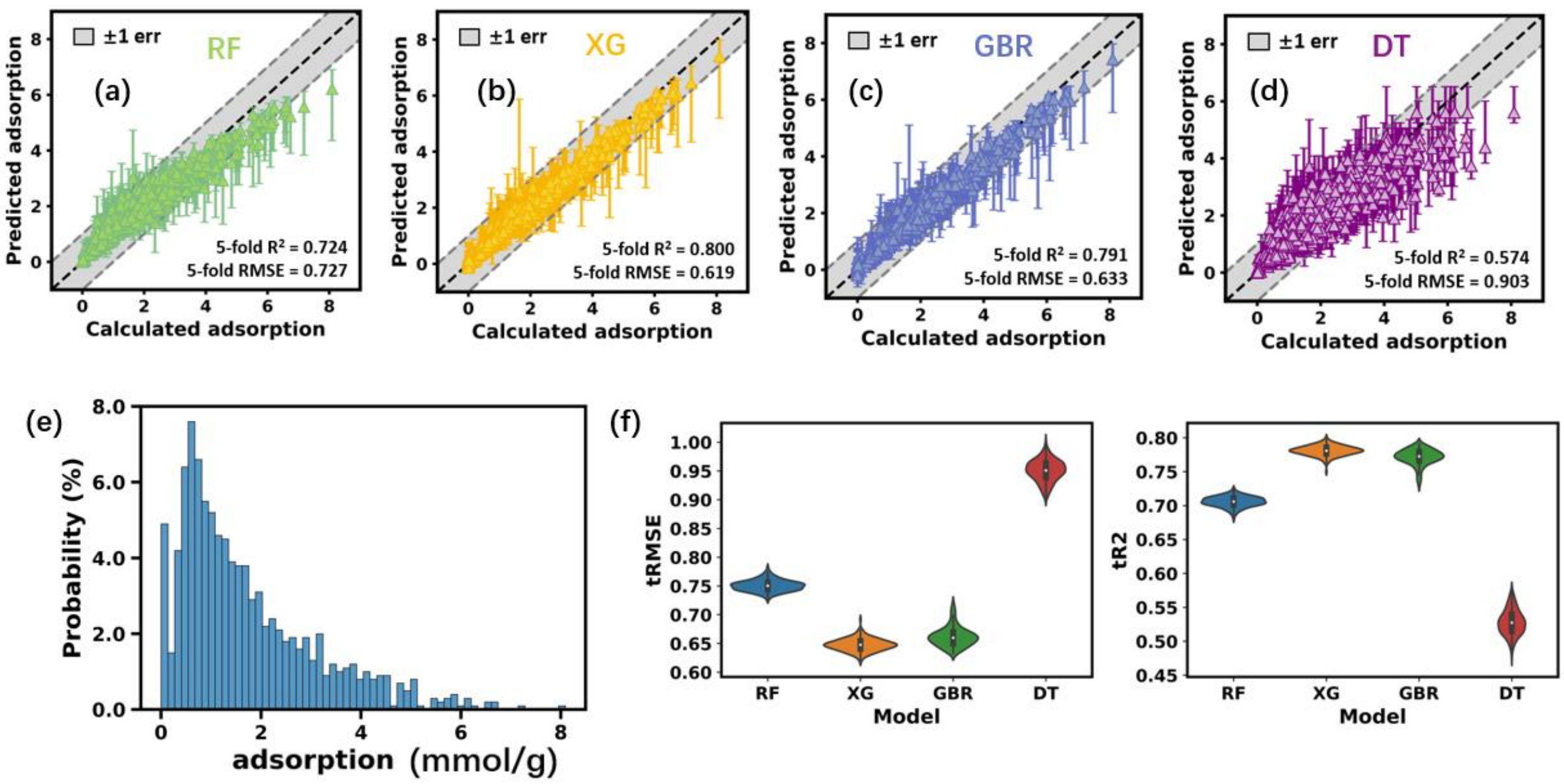

Figure 2. (a–d) Prediction results of different ML models: (a) RF model; (b) XG model; (c) GBR model; (d) DT model. (e) Numerical distribution of $CO_2$ adsorption capacity in the MOF dataset. (f) Violin plots illustrating the distribution of MSE and $R^2$ values obtained from different ML models. The $R^2$ values of the XGBoost, RF, and GBR models for predicting $CO_2$ adsorption capacities of MOFs at low pressure fall within the range of 0.7–0.8. The majority of samples have a noticeable uneven distribution and are concentrated in the lower range of 0– 2 mmol/g. Violin plots of the normalized coefficient of determination ($tR^2$) and normalized root mean square error (*tRMSE*) show that the XG model has the best generalization ability.

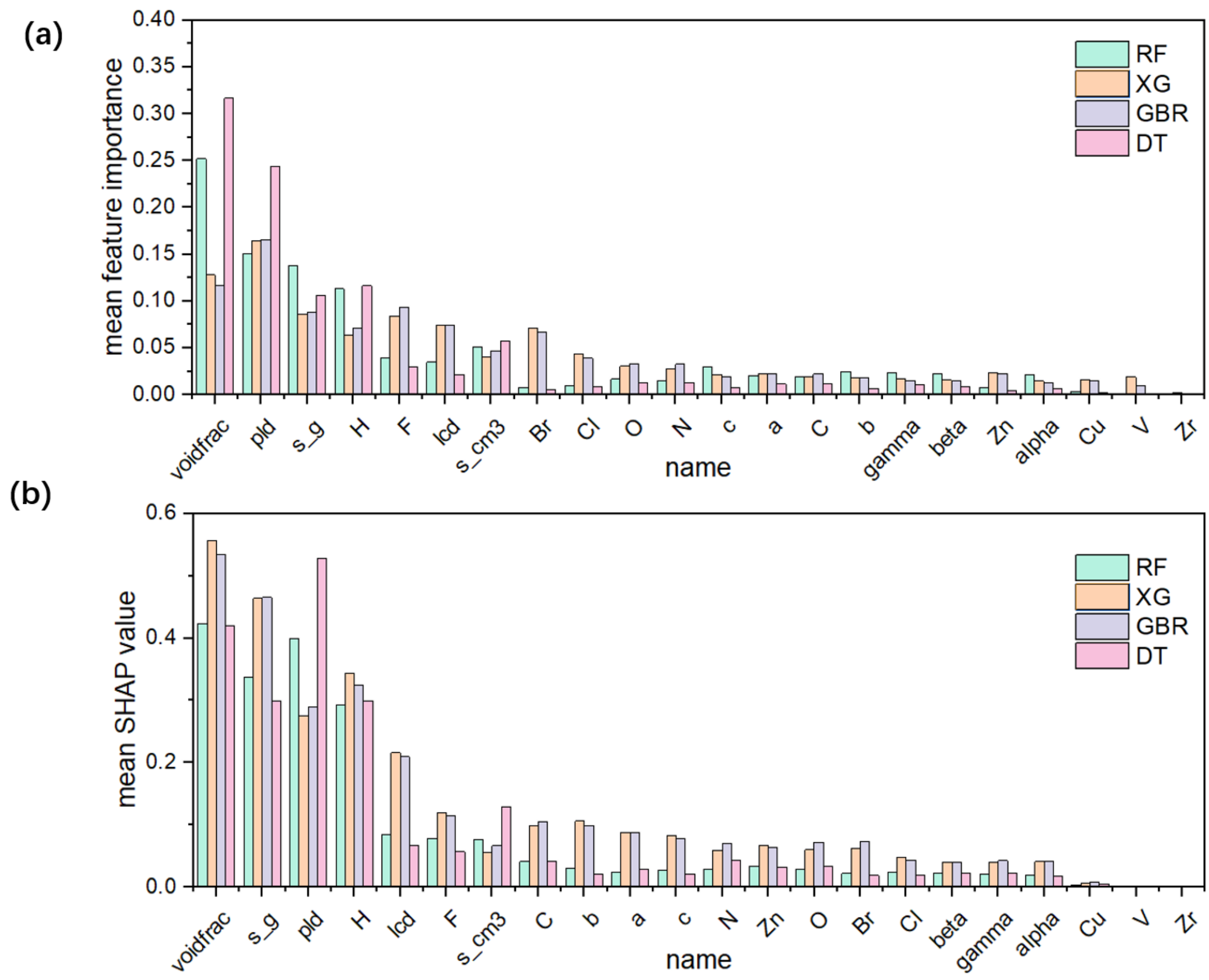

Figure 3. (a) Mean feature importance values of 22 descriptors; (b) Mean SHAP values of the 22 descriptors, obtained from the RF, XG, GBR, and DT models. The top five influential descriptors, void fraction (voidfrac), pore limiting diameter (pld), specific surface area (s_g), largest cavity diameter (lcd), and the number of hydrogen atoms (H), demonstrate that these structural features play a decisive role in determining MOF adsorption behavior for $CO_2$. Other features, such as the presence of halogens, metal content, and lattice parameters (a, b, c), also show some degree of correlation but contribute considerably less overall.

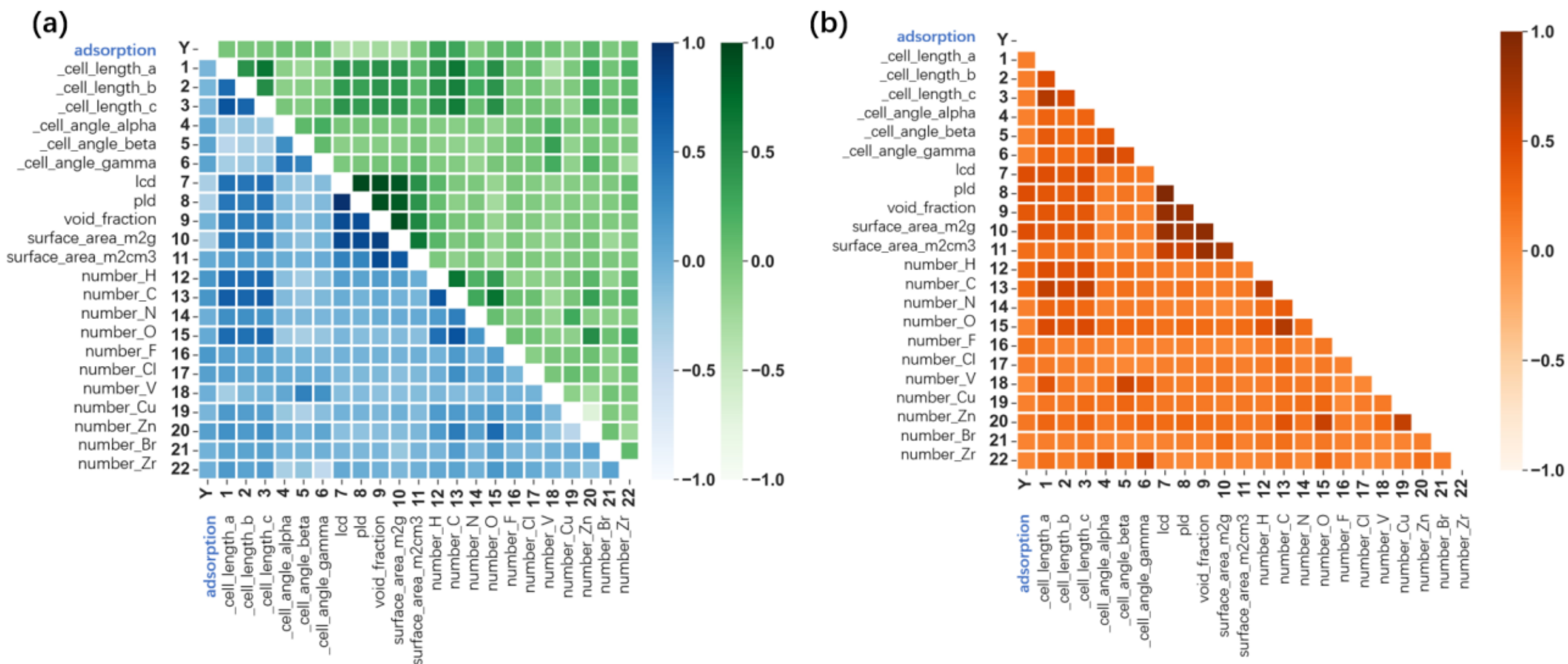

Figure 4. (a) Blue heatmap representing Pearson correlation and green heatmap representing Spearman correlation; (b) Orange heatmap illustrating distance correlation for descriptors. These three metrics describe linear, nonlinear, monotonic and non-monotonic, relationships among the features, complementing each other in identifying the key structural factors that influence $CO_2$ adsorption in MOFs.

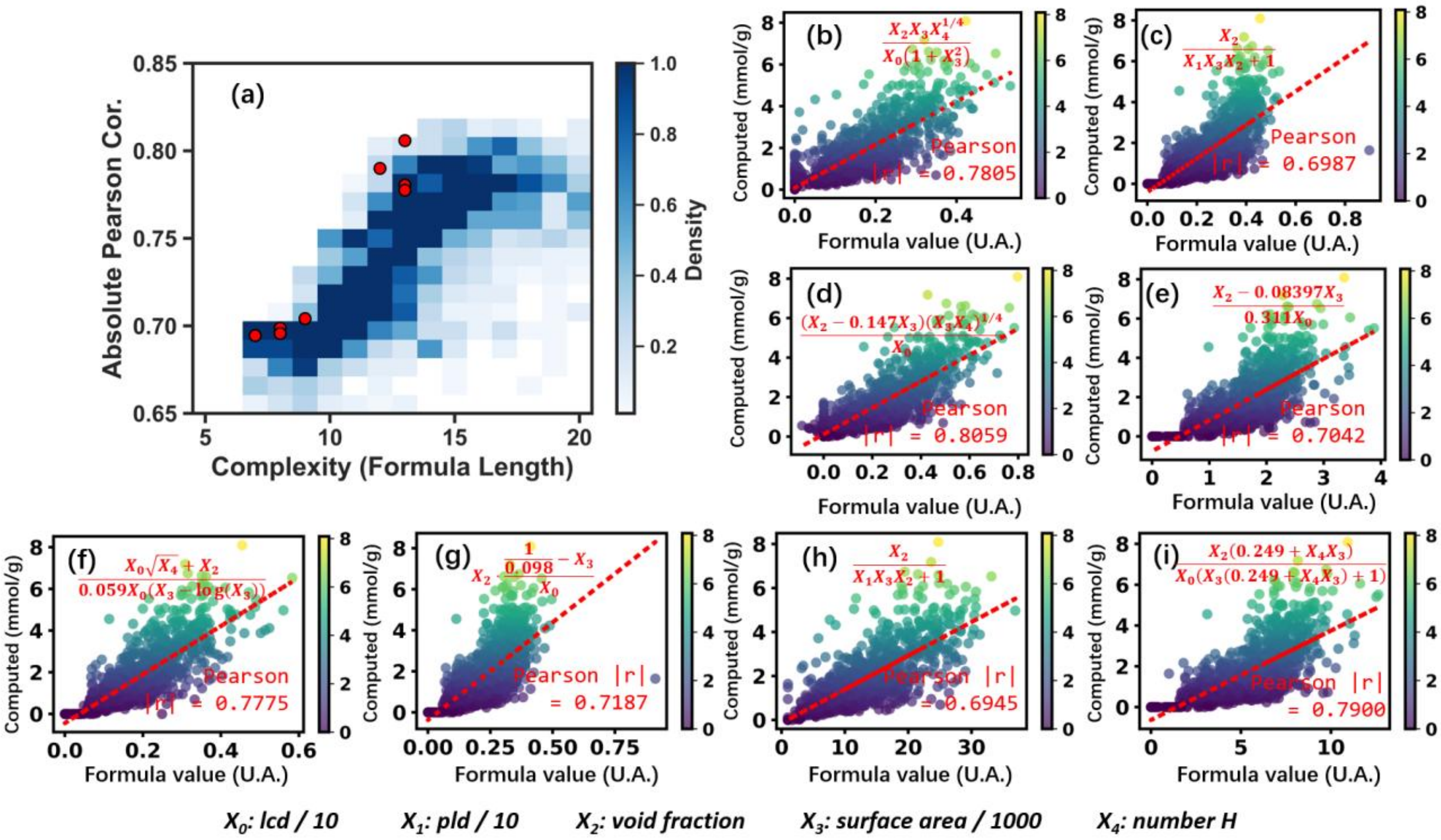

Figure 5. (a) Density plot illustrating the Pareto front between Pearson correlation coefficient (r) and the complexity of 9,073 mathematical formulas; (b–i) scatter plots comparing the $CO_2$ adsorption capacities predicted by 8 symbolic regression formulas (derived from the small training dataset) with the corresponding GCMC-computed $CO_2$ adsorption values of MOFs.

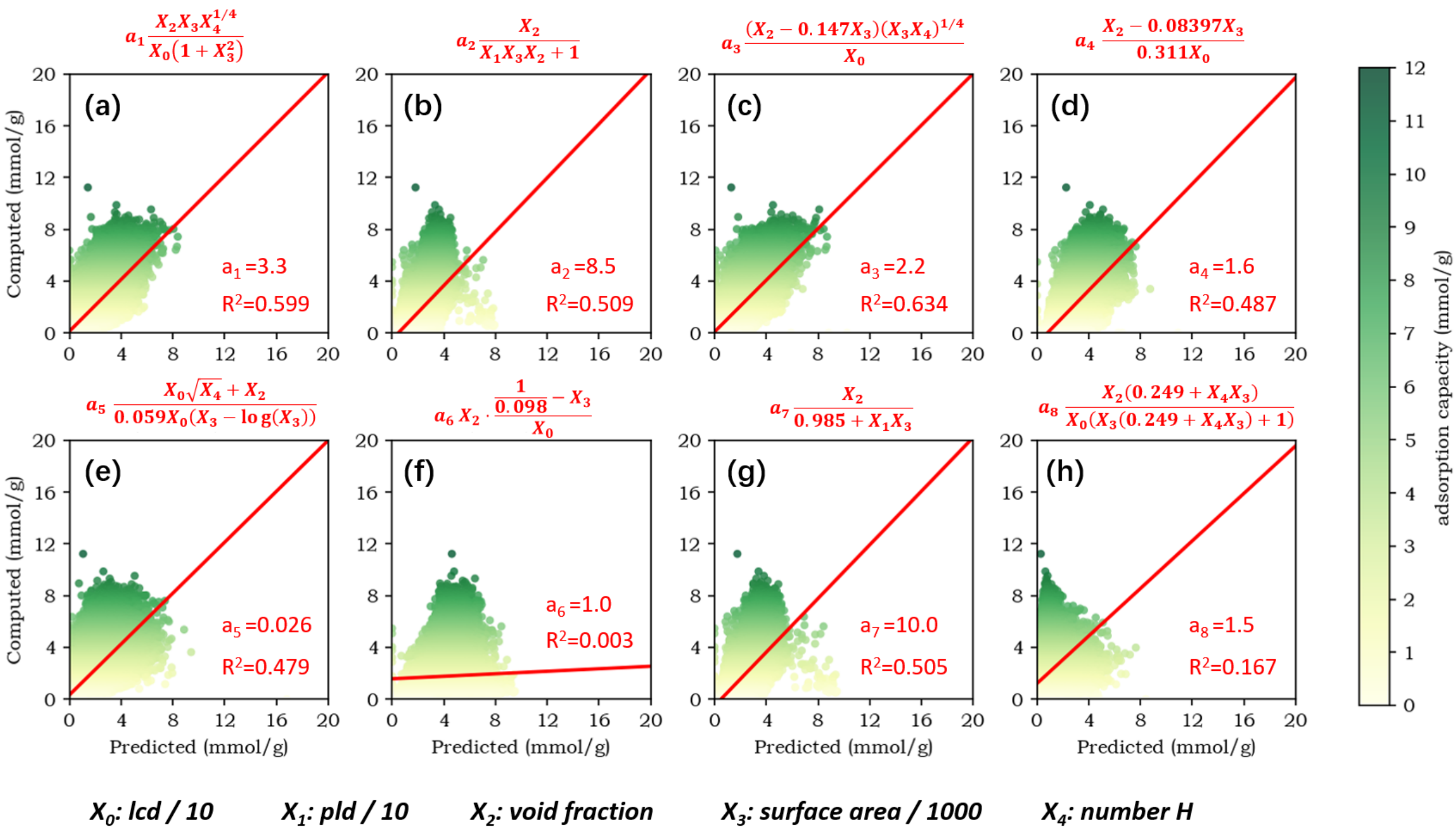

Figure 6. (a–h) Scatter plots comparing the $CO_2$ adsorption capacities of MOFs in the large-scale dataset with the predicted values obtained from 8 symbolic regression formulas.

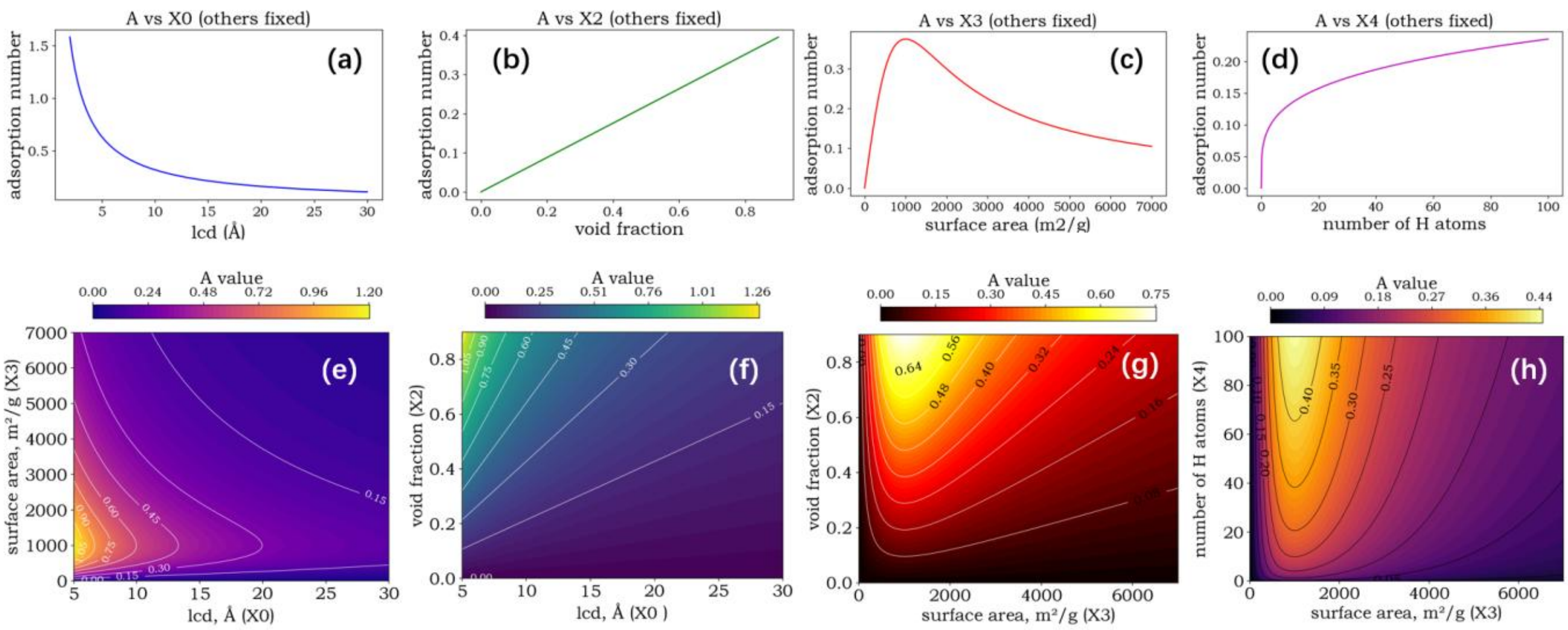

Figure 7. Single-factor analysis of the correlations between the adsorption parameter ***A*** and (a) largest cavity diameter (lcd, $X_0$), (b) void fraction, $X_2$, (c) surface area, $X_3$, and (d) number of hydrogen atoms, $X_4$. Dual-factor analysis of the correlations between ***A*** and (e) lcd, $X_0$ and surface area, $X_3$; (f) lcd, $X_0$ and void fraction, $X_2$; (g) surface area, $X_3$ and void fraction, $X_2$; and (h) surface area, $X_3$ and number of hydrogen atoms, $X_4$. All other unexamined variables are fixed at constant values for the analysis (e.g., lcd = 18 Å, void fraction = 0.45, surface area = 3500 m²/g, and number of H atoms = 50).

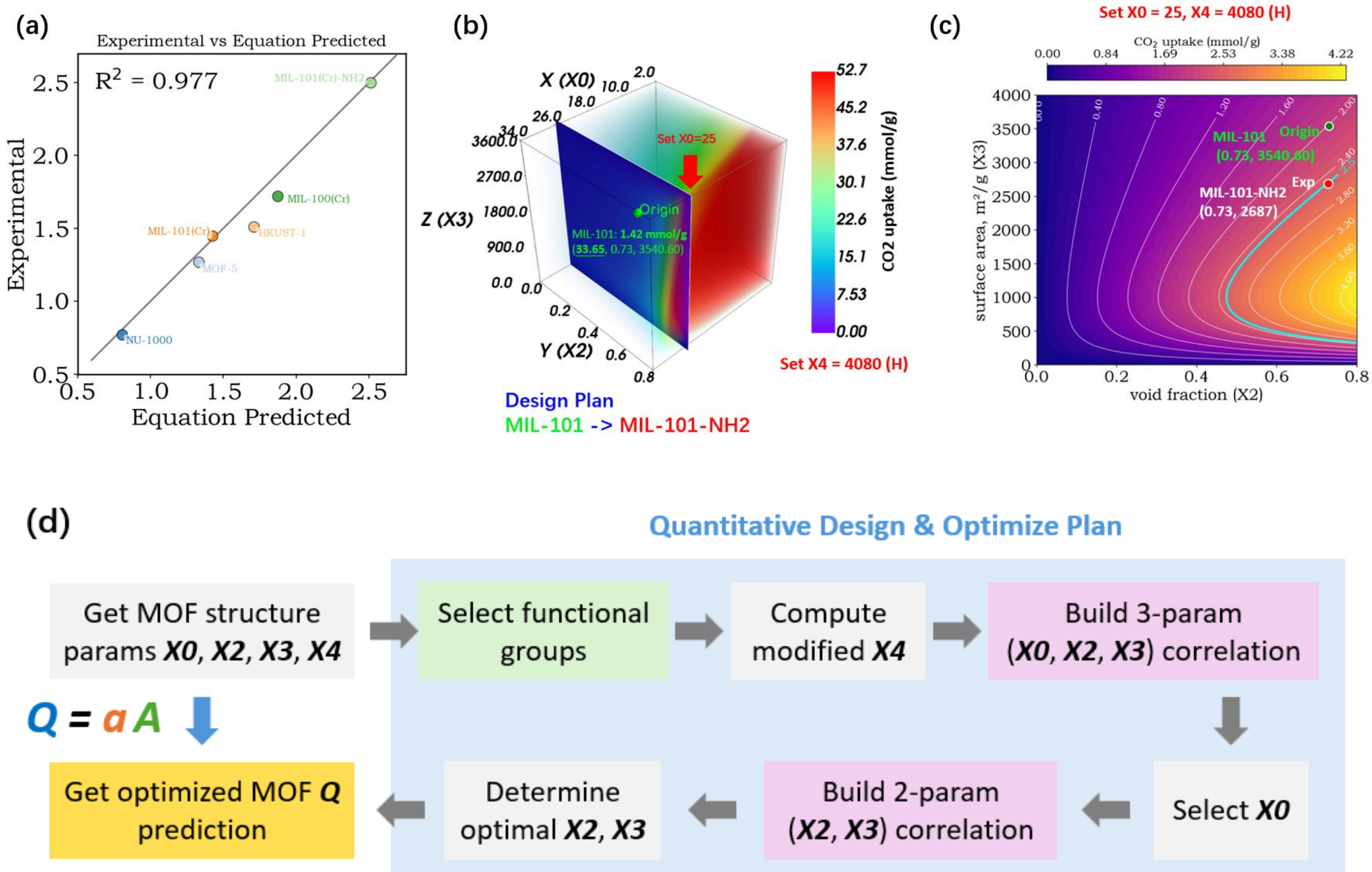

**Figure 8.** (a) Comparison between the empirical formula predictions and experimental $CO_2$ adsorption capacities of five common MOFs at 0.5 bar; (b) 3-dimensional correlation plot of $CO_2$ adsorption capacity with respect to parameters $X_0$, $X_2$, and $X_3$, based on the empirical formula under a fixed target value of $X_4$; (c) 2-dimensional heatmap showing the correlation between $CO_2$ adsorption capacity and parameters $X_2$ and $X_3$, with $X_0$ and $X_4$ held constant according to the empirical formula. (d) workflow diagram for quantitative MOF optimization and design.

**Table 1. Comparison of experimentally measured $CO_2$ adsorption capacities of MOFs at 0.5 bar with the values predicted by the empirical equation.**

| Name | lcd (Å) | pld (Å) | void frac | Sa (m2/g) | H | Exp | X0 | X1 | X2 | X3 | X4 | Eqn | [b] Eqn Accuracy |
|---|---|---|---|---|---|---|---|---|---|---|---|---|---|
| **HKUST-1**[a] | 13.19 | 6.67 | 0.68 | 2757.89 | 96 | 1.51 [62] | 1.32 | 0.67 | 0.68 | 2.76 | 96 | 1.71 | 86.91% |
| **MOF-5**[a] | 15.07 | 7.93 | 0.78 | 3740.80 | 96 | 1.27 [63] | 1.51 | 0.79 | 0.78 | 3.74 | 96 | 1.33 | 95.07% |
| **MIL-101(Cr)**[a] | 33.65 | 13.86 | 0.73 | 3540.60 | 3264 | 1.45 [64] | 3.37 | 1.39 | 0.73 | 3.54 | 3264 | 1.42 | 98.10% |
| **MIL-100(Cr)**[a] | 26.59 | 8.94 | 0.61 | 2089.29 | 1632 | 1.72 [65] | 2.66 | 0.89 | 0.61 | 2.09 | 1632 | 1.87 | 91.10% |
| **NU-1000**[a] | 29.20 | 28.31 | 0.79 | 3468.38 | 132 | 0.77 [23] | 2.92 | 2.83 | 0.79 | 3.47 | 132 | 0.81 | 95.33% |
| **ZIF-8**[a] | 11.39 | 3.41 | 0.51 | 2130.40 | 120 | 0.61 [66] | 1.14 | 0.34 | 0.51 | 2.13 | 120 | 1.90 | 31.93% |
| **UiO-66**[a] | 8.46 | 3.81 | 0.50 | 2063.13 | 96 | 1.95 [67] | 0.85 | 0.38 | 0.50 | 2.06 | 96 | 2.38 | 77.91% |
| **MIL-100(Cr)-$NH_2$**[52] | 25.00 | 13.86 | 0.73 | 2687.00 | 4080 | 2.50 [52] | 2.50 | 1.39 | 0.73 | 2.69 | 4080 | 2.51 | 99.99% |

*a*: these MOFs were all characterized using Zeo++ [68] under high-accuracy settings. Zeo++ employs Voronoi decomposition to identify probe-accessible void regions and to calculate the accessible surface area, accessible volume, Largest Cavity Diameter, and Pore Limiting Diameter.

*b*: Individual Relative Accuracy = 1 – $|y_i - \hat{y}_i| / |y_i| \times 100\%$ , where $y_i$ the true value and $\hat{y}_i$ the predicted value.

Table 2. The list of 5 descriptors.

| No | Name | Description | SR Variable |
|---|---|---|---|
| **1** | lcd | Largest Cavity Diameter (Å); 10 (Å) is the scaling constant for *lcd*; *lcd / 10* | $x_0$ |
| **2** | pld | Pore Limiting Diameter (Å); 10 (Å) is the scaling constant for *pld*; *pld / 10* | $x_1$ |
| **3** | VoidFrac | Void fraction | $x_2$ |
| **4** | S_g | Gravimetric surface area ($m^2/g$); 1000 ($m^2/g$) is the scaling constant for $S_a$; $S_a / 1000$ | $x_3$ |
| **5** | H | Number of H atoms; | $x_4$ |

Table 3. Setup of hyperparameters in the gplearn software for SR.

| Parameter | Value | Combination | Description |
|---|---|---|---|
| **Generations** | 200 | 1 | fixed at 200 evolutionary iterations |
| **Population size in every generation** | 1000 | 1 | fixed at 1000 individuals per generation |
| **Probability of crossover ($p_c$)** | [0.30, 0.90], (step = 0.05) | 746 | The Combination = 746 for $p_c$, $p_s$, $p_h$, and $p_p$ denotes the total feasible joint parameter sets explored in grid search, not 746 values for each parameter individually, under the constraint $p_c + p_s + p_h + p_p = 1$. |
| **Probability of subtree mutation ($p_s$)** | [(1-pc)/3, (1-pc)/2] (step = 0.01) | | |
| **Probability of hoist mutation ($p_h$)** | [(1-pc)/3, (1-pc)/2] (step = 0.01) | | |
| **Probability of point mutation ($p_p$)** | 1-pc-ps-ph | | |
| **Function set** | $+, -, \times, \div, \sqrt{x}, \ln x, \lvert x \rvert, -x, 1/x$ | 1 | fixed at 1 predefined operator library for all symbolic regression runs. |
| **Parsimony coefficient** | auto | 1 | fixed at automatic mode for adaptive model complexity control |
| **Metric** | Pearson Cor. | 1 | fixed at Pearson correlation coefficient as the fitness evaluation criterion. |
| **Stopping criteria** | 1.0 | 1 | fixed at a threshold value of 1.0 for terminating the evolutionary search. |
| **Random_state** | 0, 1, 2, 3, 4 | 5 | 5 independent runs using seeds 0–4 to reduce stochastic bias. |
| **Init_depth** | 2-4, 3-5, 2-6, 3-7 | 4 | 4 initialization depth ranges (2–4, 3–5, 2–6, 3–7). |

# Support Information

## Discovering Physically Interpretable Mathematical Expression for Predicting $CO_2$ Adsorption in Metal–Organic Frameworks via Machine Learning-Symbolic Regression

Yimin Shao,[1,2†] Shengluo Ma,[3†] Shenghong Ju,[3,4*] Yijun Shi,[2] Wei Li[1*]

1. Institute for Materials and Processes, School of Engineering, The University of Edinburgh, Edinburgh EH9 3FB, Scotland, UK
2. Machine Elements Division, Department of Engineering Sciences and Mathematics, Luleå University of Technology, Luleå, 97187, Sweden
3. China-UK Low Carbon College, Shanghai Jiao Tong University, Shanghai 201306, China
4. Key Laboratory for Thermal Science and Power Engineering of Ministry of Education, Department of Engineering Mechanics, Tsinghua University, Beijing 100084, China

†These authors contributed equally to this work.

*Corresponding authors. Email: wli5@ed.ac.uk , jush@tsinghua.edu.cn

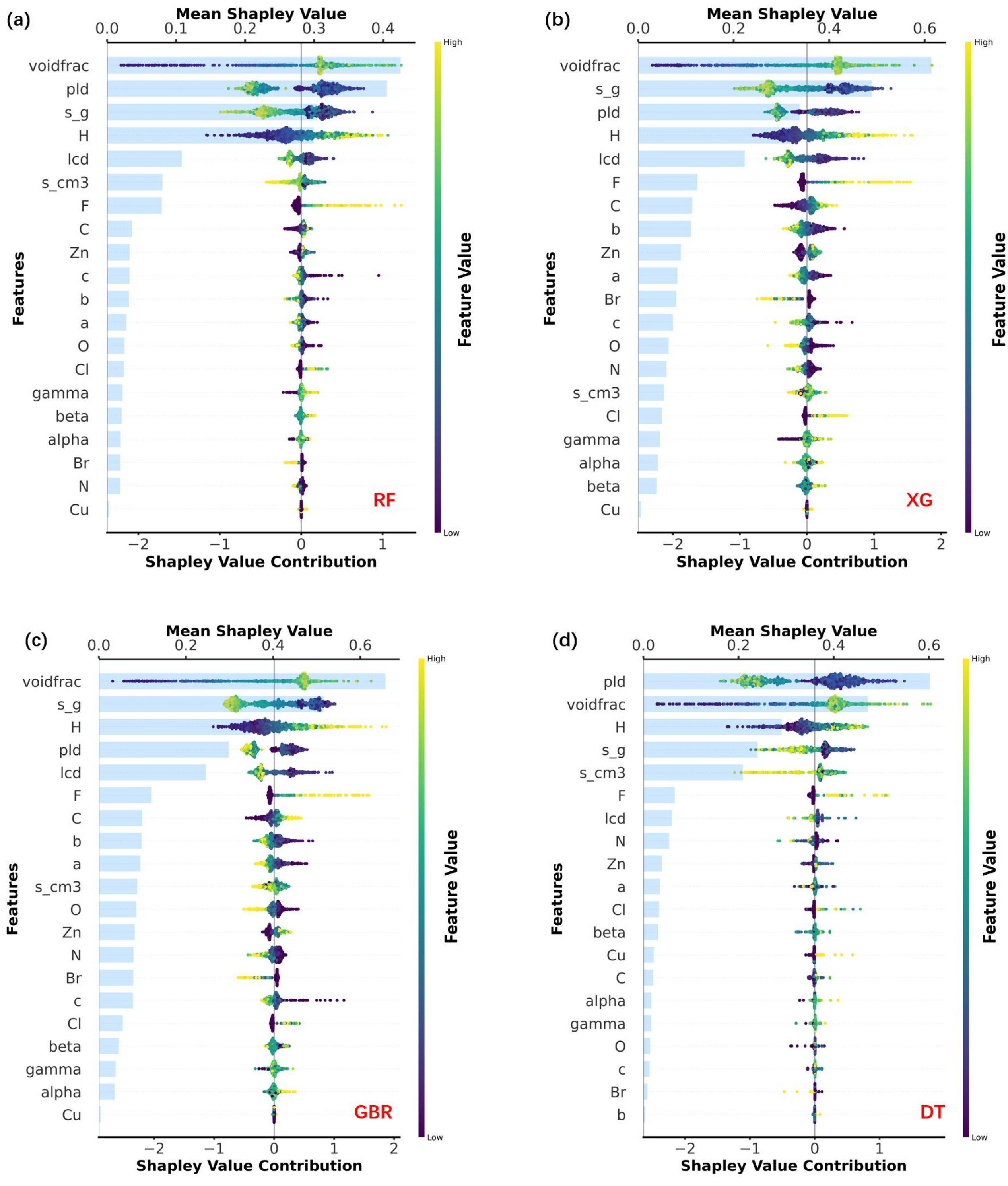

**Figure S1**. Analysis of feature importance using SHAP on (a) RF model; (b) XG model; (c) GBR model; (d) DT model; trained by SR descriptors set, including Mean absolute SHAP values for 22 descriptors and Represent the SHAP values of each descriptor related to the hMOF data in a beeswarm plot.

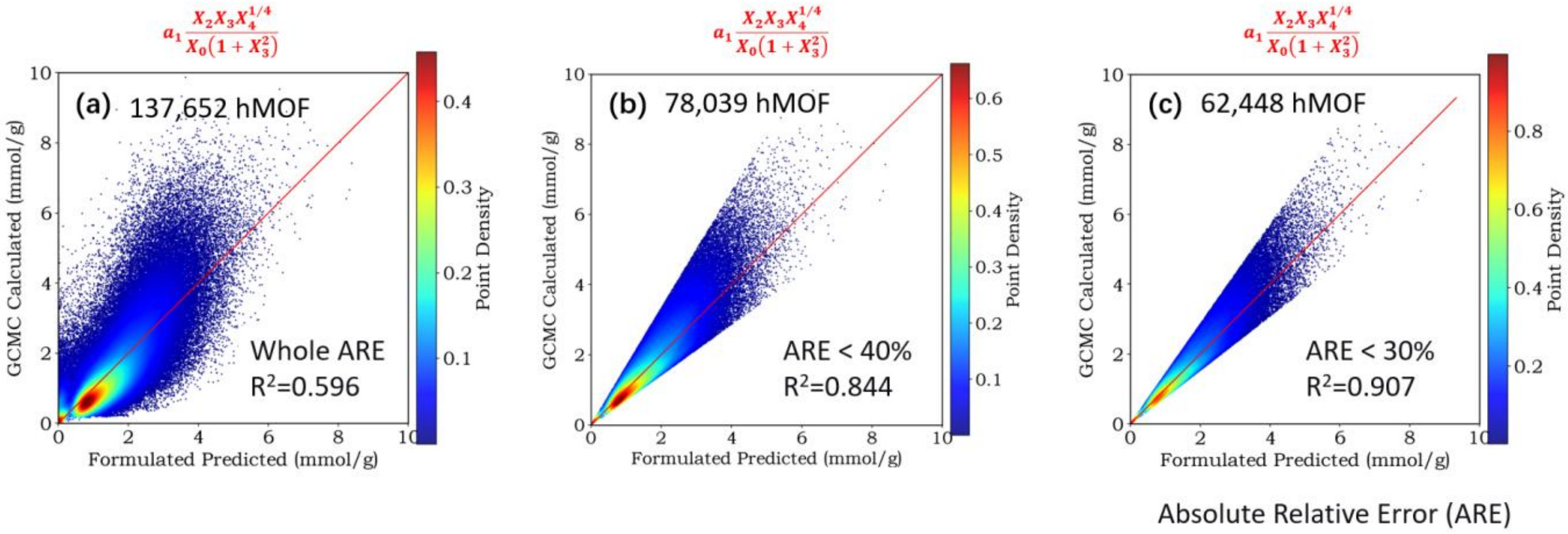

**Figure S2.** Scatter plots comparing the $CO_2$ adsorption capacities of MOFs in the large-scale dataset with the predicted values obtained from one symbolic regression formula: (a) full dataset; (b) dataset with ARE < 40%; (c) dataset with ARE < 30%.

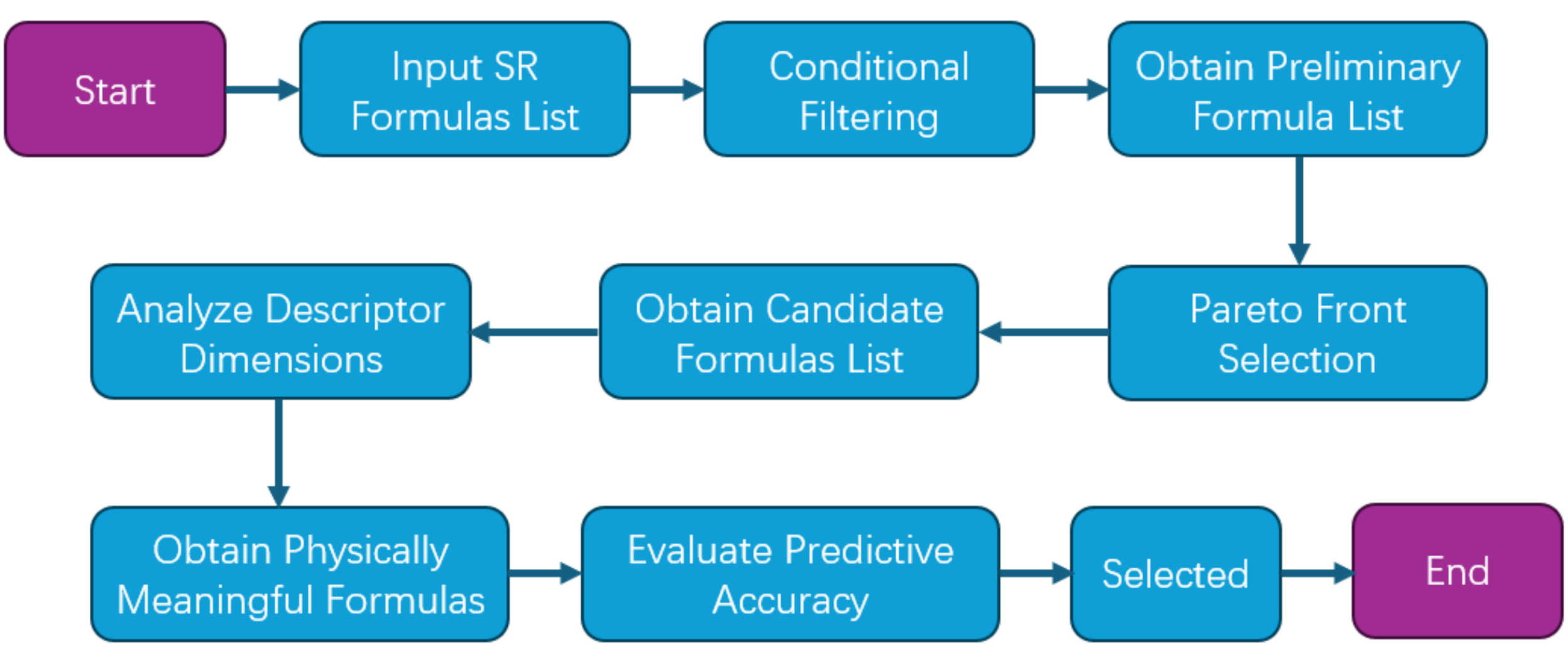

**Figure S3.** Flowchart of symbolic formula screening process.

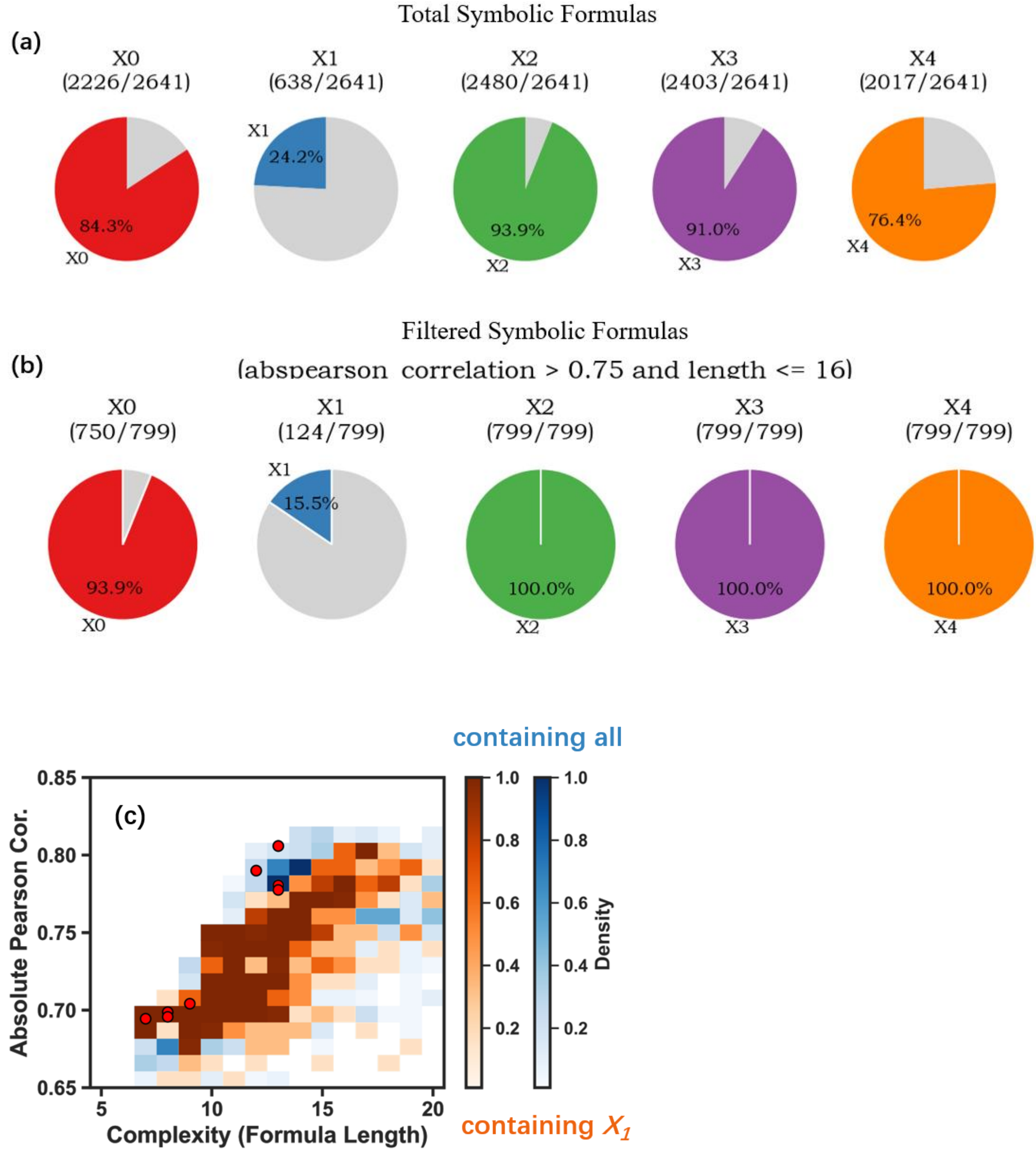

**Figure S4.** (a) Proportions of formulas containing different variables ($X_0$, $X_1$, $X_2$, $X_3$, and $X_4$) among the total 2,641 formulas; (b) proportions of formulas containing different variables ($X_0$, $X_1$, $X_2$, $X_3$, and $X_4$) among the filtered formulas; (c) comparison between the density distribution of formulas containing $X_1$(orange) and that of all variables' formulas (blue).

Table S1. The features and their description in this work.

| Feature Name | Description |
|---|---|
| **a** | cell length a |
| **b** | cell length b |
| **c** | cell length c |
| **alpha** | cell angle alpha |
| **beta** | cell angle beta |
| **Gamma** | cell angle gamma |
| **lcd** | Largest Cavity Diameter (Å) |
| **pld** | Pore Limiting Diameter (Å) |
| **voidfrac** | void fraction |
| **S_g** | gravimetric surface area (m2/g) |
| **S_cm3** | volumetric surface area (m2cm3) |
| **H** | number of H atoms |
| **C** | number of C atoms |
| **N** | number of N atoms |
| **O** | number of O atoms |
| **F** | number of F atoms |
| **Cl** | number of Cl atoms |
| **V** | number of V atoms |
| **Cu** | number of Cu atoms |
| **Zn** | number of Zn atoms |
| **Br** | number of Br atoms |
| **Zr** | number of Zr atoms |

Table S2. Results of empirical formulas' predictions for the entire hMOF dataset.

| No | Formula | MOF Data | < 30% AER | < 20 % AER | < 10 % AER |
|---|---|---|---|---|---|
| 1 | $a_1 \frac{X_2 X_3 X_4^{1/4}}{X_0(1+X_3^2)}$ | 137,652 | 62,448 (45.37 % of total) | 43,924 (31.91% of total) | 23,689 (17.21% of total) |
| 2 | $a_2 \frac{X_2}{X_1 X_3 X_2 + 1}$ | 137,652 | 51,568 (37.46 % of total) | 36,803 (26.74% of total) | 19,730 (14.33% of total) |
| 3 | $a_3 \frac{(X_2 - 0.147X_3)(X_3X_4)^{1/4}}{X_0}$ | 137,652 | 64,491 (46.85% of total) | 45,847 (33.31% of total) | 24,905 (18.09% of total) |
| 4 | $a_4 \frac{X_2 - 0.08397X_3}{0.311X_0}$ | 137,652 | 41,449 (30.11% of total) | 29,498 (21.43% of total) | 15,848 (11.51% of total) |
| 5 | $a_5 \frac{X_0\sqrt{X_4} + X_2}{0.059X_0(X_3 - \log(X_3))}$ | 137,652 | 64,149 (46.60% of total) | 44,767 (32.52% of total) | 23,281 (16.91% of total) |
| 6 | $a_6 X_2 \cdot \frac{\frac{1}{0.098} - X_3}{X_0}$ | 137,652 | 25,646 (18.63% of total) | 18,922 (13.75% of total) | 10,793 (7.84% of total) |
| 7 | $a_7 \frac{X_2}{0.985 + X_1X_3}$ | 137,652 | 50,583 (36.75% of total) | 35,914 (26.09% of total) | 19,396 (14.09% of total) |
| 8 | $a_8 \frac{X_2(0.249 + X_4X_3)}{X_0(X_3(0.249 + X_4X_3) + 1)}$ | 137,652 | 16,367 (11.89% of total) | 11,618 (8.44% of total) | 7,539 (5.48 % of total) |

% Absolute Relative Error (AER) = $|y_i - \hat{y}_i| / |y_i| \times 100\%$ , where $y_i$ the true value and $\hat{y}_i$ the predicted value.

Table S3. Physical Interpretation of Basic Mathematical Operations in Symbolic Regression Models.

| Operation | Mathematical Form | Physical Meaning |
|---|---|---|
| **Addition** | A + B | Superposition |
| **Subtraction** | A − B | Difference / Driving Force / Gradient / Δ |
| **Multiplication** | A × B | Amplification / Coupling / Scaling / Interaction |
| **Division** | A / B | Density / Normalization / Intensity |